\documentclass[sigconf]{acmart}

\acmConference[SBQS26]{Brazilian Symposium on Software Quality}{Nov 10--13, 2026}{Maringá, PR}
\setcopyright{none}
\renewcommand\footnotetextcopyrightpermission[1]{}

\usepackage[utf8]{inputenc}
\usepackage[T1]{fontenc}
\usepackage{tikz}
\usetikzlibrary{arrows.meta,positioning,calc}
\usepackage{microtype}
\hypersetup{hidelinks}

\begin{document}
\renewcommand\abstractname{ABSTRACT}
\renewcommand\keywordsname{KEYWORDS}
\renewcommand\refname{REFERENCES}
\title[Beyond Accuracy: Decomposition vs. Typed-State QA]{Beyond Accuracy: Disentangling Decomposition from Typed-State Quality Assurance in Multi-Agent LLM Pipelines}

\newif\ifanonymous
\anonymousfalse

\ifanonymous
  \author{Anonymous}
  \affiliation{%
    \institution{Anonymous Institution}
    \city{Anonymous City}
    \country{Anonymous Country}
  }
  \email{anonymous@anonymous.com}
\else
  \author{Maria Gabriela Alves Zuppardo}
  \affiliation{%
    \institution{Centro de Inform\'atica, Universidade Federal de Pernambuco (UFPE)}
    \city{Recife}
    \state{PE}
    \country{Brazil}
  }
  \email{mgaz@cin.ufpe.br}

  \author{Vinicius Cardoso Garcia}
  \affiliation{%
    \institution{Centro de Inform\'atica, Universidade Federal de Pernambuco (UFPE)}
    \city{Recife}
    \state{PE}
    \country{Brazil}
  }
  \email{vcg@cin.ufpe.br}

  \renewcommand{\shortauthors}{Zuppardo et al.}
\fi

\begin{abstract}
Multi-agent LLM pipelines are often credited with gains in software engineering tasks, but it is
rarely clear whether a gain stems from the typed contract between agents or from the specialized
prompts that decomposition enables. We study this attribution problem in Requirements Engineering,
comparing three architectures for classifying requirements as functional or non-functional and then
prioritizing them with MoSCoW: a single-call baseline; a two-call variant that reuses the pipeline's
specialized prompts but passes an unvalidated intermediate; and a pipeline with a
Pydantic-validated typed state. The 18 conditions cover PROMISE NFR+ (625 requirements), three
locally hosted small language models, and two languages. The F1 gain over the baseline is
associated with the two-call, specialized-prompt variant (up to $+0.075$), whereas the typed state
adds a small difference ($-0.003$ to $+0.013$) that is significant in one of six comparisons.
Beyond accuracy, the typed pipeline contains out-of-range model outputs at the agent boundary,
while the unvalidated variant fails later and drops the requirement, and it shows a small,
reproducible increase in inter-model MoSCoW consistency whose cause our design cannot isolate.
The study is associative: the contrasts bundle prompts, call count, and staged context, and the
results come from three 7--8B models on one dataset. The implementation and replication package
are archived on Zenodo (\url{https://doi.org/10.5281/zenodo.23380444}).
\end{abstract}


\maketitle

\section{Introduction}
\label{sec:intro}

Large Language Models (LLMs) are increasingly deployed as components of
larger software pipelines, where their outputs feed downstream decisions
with limited human review~\cite{liu2024agentse}. This raises a quality
assurance question that has received little controlled empirical
attention in the Brazilian Software Quality community: when a pipeline is
decomposed into specialised agents connected by a typed contract, what
does that contract buy beyond the accuracy of the underlying model calls?
Recent SBQS work shares the evaluate-beyond-accuracy concern in adjacent
settings, such as judging LLM-generated unit tests against
mutation-testing criteria~\cite{cabral2025mutation} and assessing
generative AI as a substitute for usability
inspectors~\cite{campos2025inspectors}. Practitioners likewise cite
reliability and weak contextual grounding as primary barriers to adopting
LLMs in Requirements Engineering~\cite{santos2025adoption}.

None of the studies we surveyed isolates the contribution of an
architectural mechanism, such as a typed inter-agent contract, from that
of prompt design. When a pipeline outperforms a single-agent baseline,
the gain is often read as evidence that the contract improves output
quality; whether prompting alone would produce the same gain is rarely
checked, and systematic reviews confirm that this conflation remains
unresolved~\cite{zadenoori2025re,hemmat2025research}
(\S\ref{sec:related}). F1 is not this paper's object of interest but the
instrument we use to rule out one explanation before asking a more
specific question: we show that the typed contract barely moves F1
($\Delta_{\text{s}} \approx 0$), which points to decomposition and its
task-specific prompts as the accuracy driver and leaves open what
differences accompany the typed contract, a question F1 cannot answer
(\S\ref{sec:results:rq2}, \S\ref{sec:results:errorprop}).

We therefore ask not only whether a multi-agent pipeline refines
classification accuracy, but \emph{which} architectural element drives
any gain and what differences accompany the typed contract. We
operationalise \emph{quality assurance} narrowly, as early structural
validation and cross-model behavioural consistency, loosely related to
ISO/IEC~25010's \textit{Reliability} (we do not measure
\textit{Maintainability}) rather than the \textit{Functional Suitability}
that F1 already captures. We
compare three architectures on FR/NFR classification followed by MoSCoW
prioritisation, using PROMISE
NFR+~\cite{silva2024promiseplus_zenodo,silva2024promiseplus_sbes} (625
requirements): a \textbf{baseline} that issues a single LLM call per
requirement; a \textbf{pipeline} of two specialised agents orchestrated
with LangGraph~\cite{langgraph2024} over a Pydantic-validated typed
state; and a \textbf{two-call baseline} that reuses the pipeline's
task-specific prompts but passes an unvalidated intermediate object.
Because the second call also receives the first call's decided type and
category, the baseline-to-two-call difference is a \emph{decomposition
effect} that bundles prompt wording, call count, and staged context
(\S\ref{sec:architecture:limits}, \S\ref{sec:results:rq2}). All three run
on the same three locally served SLMs and decoding settings, yielding an
18-condition grid (3 models $\times$ 3 architectures $\times$ 2
languages).

Here, ``disentangling'' refers to comparing architectural contrasts; it does not imply causal
isolation of the typed contract's effect.

\textbf{Contributions.}
(1)~A controlled empirical study (18 conditions, $n = 625$; paired
McNemar and Wilcoxon tests, Cohen's $h$ and paired $g$, Bonferroni;
Fleiss' $\kappa$ with a paired bootstrap; and a sensitivity analysis
retaining parse-failure fallbacks) that separates the accuracy effect of
decomposition from the quality-assurance properties that accompany a
typed contract, corroborated by a direct baseline-vs-two-call comparison
(\S\ref{sec:results:rq2}).
(2)~Evidence that a near-zero F1 difference does not make the typed
contract dispensable, restricted to two properties: containment of
out-of-range fields at the agent boundary (classification faults one
call earlier than the unvalidated ablation), verified by fault injection
(\S\ref{sec:results:errorprop}), and a small, statistically confirmed
increase in inter-model MoSCoW consistency (\S\ref{sec:results:moscow}).
Confidence-aware routing, the other mechanism the design attributes to
the typed state, fired for only 3 of 3{,}750 pipeline predictions (all
parse fallbacks), so it is validated only through the synthetic
fault-injection test, not on the PROMISE NFR+ grid.
(3)~An open-source pipeline built on a reusable
\texttt{BaseAgent[Item, Output]} abstraction.
(4)~A replication package (prompts, runner, manifests, raw predictions
for all 36 executions, analysis scripts, tests) archived on Zenodo
(DOI~\texttt{10.5281/zenodo.23380444}).

\section{Background}
\label{sec:background}

\textbf{LLM-based agents.} LLMs have evolved from passive text generators
into agents that invoke tools, maintain state, and work with other
agents~\cite{liu2024agentse}. Multi-agent systems compose such agents
into pipelines in which each handles a sub-task and communicates through
typed messages or shared state, reflecting classical design principles
(task specialisation, typed contracts, composability). Their empirical
benefit remains open, since shared context may offset the gains of
decomposition, and it splits into two facets that this paper treats
separately: \emph{functional accuracy}, and \emph{quality attributes}
(e.g., ISO/IEC~25010 \textit{Reliability}) that an
accuracy metric does not capture.

\textbf{RE and NFR classification.} PROMISE
NFR+~\cite{silva2024promiseplus_zenodo,silva2024promiseplus_sbes}
provides 625 requirements across one functional class and ten NFR
categories, the canonical dataset for controlled classification studies
in the Brazilian community~\cite{martin2025security}. MoSCoW
prioritisation~\cite{clegg1994casemethod} (Must, Should, Could, Won't
Have) lacks a public ground truth, which motivates treating MoSCoW
consistency as a quality proxy rather than an accuracy metric
(\S\ref{sec:results:moscow}).

\textbf{Small language models.} SLMs ($\leq$ 8B parameters) are
competitive on SE classification
tasks~\cite{rejithkumar2025nice,zadenoori2025size}, and local deployment
enables controlled, versioned experiments. Structured-output compliance,
however, varies across SLMs~\cite{huang2025hallucination}, a failure mode
that matters inside pipelines, where a malformed output at one agent's
boundary can spread into the next agent's context
(\S\ref{sec:results:errorprop}).

\section{Related Work}
\label{sec:related}

Recent work has investigated the use of Large Language Models (LLMs) in Requirements Engineering (RE) for tasks such as the classification of functional and non-functional requirements (FR/NFR), prioritization, traceability, and requirements generation. However, most studies evaluate the models as standalone predictors and emphasize classification accuracy, paying less attention to the architectural mechanisms that coordinate multiple interactions between LLMs.
FR/NFR classification has been extensively investigated using the PROMISE NFR+ benchmark~\cite{silva2024promiseplus_zenodo,silva2024promiseplus_sbes}. Recent studies compare proprietary and open-source models under various prompting strategies~\cite{martin2025security}, combine fine-tuned Small Language Models (SLMs) with explainability techniques~\cite{rejithkumar2025nice}, and analyze performance differences between compact and proprietary models~\cite{zadenoori2025size}. In general, these works aim to improve predictive performance while keeping the underlying processing architecture unchanged.
From a broader perspective, systematic reviews by Zadenoori et al.~\cite{zadenoori2025re} and Hemmat et al.~\cite{hemmat2025research} identify growing interest in the use of LLMs in RE but also point out a lack of rigorous, reproducible evaluations of architectural design decisions. Concurrently, LLM-based multi-agent systems are gaining traction in software engineering tasks, driven by the decomposition of work among specialized agents and explicit orchestration mechanisms~\cite{liu2024agentse}. Although existing studies often report end-to-end performance gains, few isolate the effects of decomposition and the architectural decisions governing agent interaction.
This work addresses this attribution problem through a controlled experimental design. We compare three configurations: a single-call baseline, a multi-agent pipeline with typed contracts between agents, and a two-call baseline without typed state. By keeping the underlying models and decoding configurations constant, the design makes it possible to distinguish the effects of task-specific decomposition and prompting from those associated with the typed contract between agents. The
gap we claim is relative to the two systematic reviews above and to
SBQS/SBES proceedings of 2024--2025; our survey was not systematic.

\section{Implementation and Experimental Platform}
\label{sec:architecture}

All three architectures address the same compound RE task, FR/NFR
classification followed by MoSCoW prioritisation, and share the dataset
adapter, model factory, decoding parameters, retry policy, concurrency
level, and metric computation. As Table~\ref{tab:arch} shows, they differ
in more than the intermediate representation; we state each difference so
that effects are not attributed to a single factor
(\S\ref{sec:architecture:limits}).

\subsection{Architectural Overview}
\label{sec:architecture:overview}

The \textbf{baseline} classifies and prioritises in a single LLM call
(\texttt{BaselineAgent}, joint prompt, no intermediate state). The
\textbf{two-call baseline} (\texttt{TwoCallBaselineAgent}) uses the same
task-specific prompt builders as the pipeline but passes the intermediate
result in an unvalidated Python dataclass, so Pydantic validation happens
only when the final output is assembled. The \textbf{pipeline} divides the
work between two agents that exchange a validated Pydantic object,
orchestrated by LangGraph. In both two-call arms the prioritisation prompt
receives the type and category decided by the first call.

Comparing the baseline with the two-call baseline bundles decomposition,
call count, task-specific prompts, and staged context. Comparing the
two-call baseline with the pipeline bundles the typed contract, the
validation boundary, the handling of confidence, and LangGraph
orchestration. Neither comparison isolates a single component.

\begin{table*}[!t]
\caption{Differences between the three architectures. Shared by all: model, temperature, workers, retry policy, dataset adapter, response parsing, and metrics. EN conditions use the original English text and an English prompt; PT conditions use the translated text and a Portuguese prompt.}
\label{tab:arch}
\centering
\small
\begin{tabular}{@{}p{4.2cm}p{3.2cm}p{4.2cm}p{5.0cm}@{}}
\toprule
\textbf{Property} & \textbf{Baseline} & \textbf{Two-call} & \textbf{Pipeline} \\
\midrule
LLM calls per requirement & 1 & 2 & 2 \\
Prompts & joint & task-specific & task-specific (same as two-call) \\
Intermediate result & none & plain dataclass & validated Pydantic object \\
First Pydantic validation & after the single call & after the second call & after the first call (agent boundary) \\
Confidence given to the prioritiser & --- & fixed at 1.0 & passed; below 0.70 the prompt asks for conservative labels \\
Orchestration & direct agent call & direct agent call & LangGraph \texttt{StateGraph} \\
\bottomrule
\end{tabular}
\end{table*}

\subsection{Design and Pipeline}
\label{sec:architecture:design}
\label{sec:architecture:baseline}
\label{sec:architecture:pipeline}

The code follows a layered structure with a one-way dependency rule
(presentation, application, domain, infrastructure; the domain layer
imports from none of the others). Four patterns carry the experimental
control: an \emph{Abstract Factory} (\texttt{build\_llm}) selects the
provider by model-name prefix; an \emph{Adapter}
(\texttt{PromiseAdapter}) isolates the dataset schema from the prompts; a
\emph{Template Method} (\texttt{BaseAgent[Item, Output]}) fixes the agent
contract; and a \emph{Strategy} (\texttt{OrchestrationStrategy}) lets one
runner execute every architecture from the same configuration. Prompts
are versioned files whose version is recorded in each manifest.

The pipeline consists of \texttt{ClassificationAgent} and
\texttt{PrioritizationAgent}. Agents process requirements in a pool of
three workers, retry each call up to three times with exponential
back-off, and preserve input order; the global rank is recomputed by
sorting on \texttt{priority\_score}. The prioritisation prompt receives
the classifier's confidence and, below $0.70$, instructs the model to be
conservative; the threshold is a design choice, not a calibrated
parameter. The workflow is a LangGraph \texttt{StateGraph} with the fixed
topology $\texttt{classifier}\rightarrow\texttt{prioritizer}\rightarrow
\texttt{END}$; only the pipeline uses LangGraph.

\label{sec:architecture:state}
The agents share a \texttt{PipelineState} (a Pydantic \texttt{BaseModel})
whose requirement entity grows by inheritance
(\texttt{Requirement}~$\rightarrow$~\texttt{ClassifiedRequirement}~$\rightarrow$
\texttt{PrioritizedRequirement}). Agent outputs have bounded fields:
\texttt{confidence} and \texttt{priority\_score} lie in $[0,1]$,
\texttt{priority\_rank} is at least 1, and the requirement type and MoSCoW
label are enumerations. A value outside these bounds raises a
\texttt{ValidationError} when the typed object is built, inside the agent
and immediately after the call; validation is thus a property of these
constructors, not of LangGraph's state assignment. The agent's parser
catches the error and substitutes a flagged fallback
(\S\ref{sec:architecture:robust}).

\subsection{Robustness Layer}
\label{sec:architecture:robust}

Small models frequently return malformed or off-schema JSON. Every agent
extracts the outermost JSON object with a regular expression (duplicated
in each agent rather than centralised). The classification-producing
agents additionally normalise the category (mapping Portuguese or English
prose to the PROMISE code) and repair a recurrent off-schema pattern in
which the model returns an NFR category code (e.g., \texttt{PE}) in the
requirement-type field, reading it as NF. The repair was needed only by
\texttt{mistral:7b}, in 16 of 625 PT and 7 of 625 EN outputs, with
identical counts in the pipeline classifier and the two-call baseline. If
parsing or validation still fails, the agent returns a deterministic
fallback instead of raising: \texttt{FUNCTIONAL} with confidence zero for
classification and \texttt{COULD\_HAVE} with score 0.5 for
prioritisation, so a batch is never aborted. Fallbacks are identifiable
(confidence zero and an error string) and are the outputs excluded or
retained in the sensitivity analysis (\S\ref{sec:methodology:metrics}).

\subsection{Execution, Traceability, and Verification}
\label{sec:architecture:harness}
\label{sec:architecture:verification}

One \texttt{ExperimentRunner} executes every architecture from a frozen
configuration and writes a manifest (strategy, model, language, seed,
temperature, prompt version, dataset path and MD5 hash, LangGraph
version, elapsed time, code commit) and a results file with the metrics
and every prediction. Per-call JSONL traces were retained only for the
two-call supplement runs. The main grid has \textbf{12 conditions}: six
for the baseline and six for the pipeline (three models, two languages);
the two-call ablation adds six, for \textbf{18 in total}.

The unit tests cover agent parsing, category normalisation and fallbacks,
the domain models, the runner and manifest, the statistical routines, the
language rule (a test fails if any agent sends Portuguese text in an
English condition), and the confidence rule (the pipeline's prioritiser
receives the classifier's confidence; the two-call baseline fixes it at
1.0). A fault-injection campaign injects 13 faults (six value-range
violations, e.g.\ \texttt{confidence} $1.4$ or \texttt{priority\_score}
$-0.1$, and seven parse-level faults); the tests assert that the typed
pipeline contains each value-range violation at the agent boundary,
whereas the unvalidated intermediate fails only when the output model is
assembled (\S\ref{sec:results:errorprop}).

\subsection{Known Limitations of the Implementation}
\label{sec:architecture:limits}

Five aspects bound what the experiments can attribute. First,
confidence-aware routing is wired but almost never fires: only three of
3,750 pipeline predictions (two for \texttt{qwen2.5:7b} PT, one for
\texttt{mistral:7b} EN, all parse fallbacks excluded from the metrics)
have confidence below $0.70$, so its isolated effect cannot be evaluated;
the two-call baseline fixes the confidence at 1.0. Second, response
parsing is regex-based and duplicated across agents, so a response with
several brace-delimited blocks fails extraction and becomes a fallback.
Third, \texttt{seed} governs dataset sampling only and is not forwarded
to the model server, so determinism rests on temperature~0. Fourth,
agents write their results into the shared state object in place. Fifth,
we evaluated the design only in RE, although we expect the principles to
apply to other multi-stage SE tasks.

\section{Methodology}
\label{sec:methodology}

\subsection{Research Questions}
\label{sec:methodology:rq}

Two research questions guide this study, complemented by a
language-effect analysis (\S\ref{sec:results:language}) that is reported
alongside the RQs but is not itself a formal RQ, since it varies text
language and prompt language jointly rather than isolating either as a
controlled factor.

\textbf{RQ1.} To what extent does the pipeline outperform the
single-agent baseline in FR/NFR binary classification? We report Macro F1
and test the paired per-requirement outcomes across the six
model--language combinations with the Wilcoxon signed-rank test
(confidence-zero fallbacks excluded) and the exact McNemar test
(fallbacks retained), with Bonferroni correction
($\alpha' = 0.05/6 = 0.0083$).

\textbf{RQ2 (central).} To what extent does decomposition with its
task-specific prompts, rather than the typed inter-agent state contract,
account for the F1 difference between the pipeline and the single-agent
baseline, and what differences accompany the typed state contract that F1
does not capture? We examine this through two architectural contrasts,
$\Delta_{\text{p}} = \text{F1(two-call)} - \text{F1(baseline)}$ and
$\Delta_{\text{s}} = \text{F1(pipeline)} - \text{F1(two-call)}$. The
first bundles task-specific prompts, an additional call, and staged
context. The second bundles the typed Pydantic state, its validation
boundary, confidence-conditioned prompting (which fired for only 3 of
3,750 pipeline predictions, \S\ref{sec:architecture:limits}), and
LangGraph orchestration. Neither contrast isolates a single mechanism,
and because the unified baseline prompt and the task-specific prompts
cannot be made strictly identical (\S\ref{sec:discussion:internal}),
$\Delta_{\text{s}}$ is read as an upper bound on the F1 difference
associated with the typed state rather than as a point estimate.
Validation and error-propagation behaviour are examined separately
(\S\ref{sec:results:rq2}, \S\ref{sec:results:errorprop}).

Inter-model MoSCoW consistency is reported as a software quality
observation (Fleiss' $\kappa$ with a paired bootstrap of
$\Delta\kappa$; \S\ref{sec:results:moscow}) rather than as accuracy,
because PROMISE NFR+ provides no priority ground truth.

\textbf{Complementary analysis: language effect.} We compare an English
condition (original text and English prompt) with a Portuguese condition
(translated text and Portuguese prompt) across all three architectures.
Because text and prompt language change together, the results capture
their joint effect rather than either in isolation; the residual
confound is acknowledged in \S\ref{sec:discussion:external}
(\S\ref{sec:results:language}).

\subsection{Dataset}
\label{sec:methodology:dataset}

We use PROMISE
NFR+~\cite{silva2024promiseplus_zenodo,silva2024promiseplus_sbes},
the canonical benchmark for requirements classification, comprising
625 requirements: 255 functional (F) and 370 non-functional (NF)
across 10 categories (A, FT, LF, MN, O, PE, PO, SC, SE, US).

The original dataset is in English. To support the language-effect
analysis, we produced a Portuguese variant by machine-translating each
requirement with \texttt{deep\_translator} (Google Translate, 2024).
Both variants share identical ground-truth labels. We checked
translation quality via back-translation (PT$\to$EN, BERTScore F1 with
\texttt{roberta-large}): the mean score was 0.9646 across all 625
pairs, and 624 of 625 scored $\geq$ 0.90. The single below-threshold
item was manually inspected and confirmed semantically equivalent.

\subsection{Architectural Variants}
\label{sec:methodology:variants}

The architectural variant is the primary independent variable for
RQ1: baseline, pipeline, and two-call baseline (\S\ref{sec:architecture})
all accept the same 625 requirements and produce outputs of identical
structure, so that architecture is the main manipulated factor (the
remaining implementation differences are listed in
Table~\ref{tab:arch} and \S\ref{sec:architecture:limits}). Their
mechanics are described in \S\ref{sec:architecture:overview}--\ref{sec:architecture:baseline};
in brief, the baseline issues one unified LLM call, the pipeline
decomposes the task via a typed \texttt{PipelineState}, and the
two-call baseline reuses the pipeline's prompts without the typed
contract.

\subsection{Models and Configuration}
\label{sec:methodology:models}

We evaluate three SLMs served locally via Ollama 0.24.0
(Table~\ref{tab:models}). Local deployment removes network and API
dependencies, although decoding at temperature~0 is not bit-deterministic
(\S\ref{sec:discussion:internal}). Prior work
reports that SLMs achieve approximately 2\,pp lower F1 than large
proprietary models on PROMISE-family
datasets~\cite{rejithkumar2025nice,zadenoori2025size}.

\begin{table}[!ht]
\caption{Models evaluated. All use Q4\_K\_M quantization.}
\label{tab:models}
\begin{tabular}{lll}
\toprule
\textbf{Model} & \textbf{Params} & \textbf{Quant.} \\
\midrule
qwen2.5:7b  & 7.6B & Q4\_K\_M \\
llama3.1:8b & 8.0B & Q4\_K\_M \\
mistral:7b  & 7.2B & Q4\_K\_M \\
\bottomrule
\end{tabular}
\end{table}

Fixed hyperparameters: temperature $= 0.0$ (greedy decoding); seed
$= 42$ fixes the dataset sample only and is not forwarded to the model server. All experiments were run on a single machine (Intel Core
i5-13450HX, 16 GB RAM, NVIDIA RTX 3050 6 GB). Each agent processes
requirements in parallel (\texttt{max\_workers} = 3) with
exponential-backoff retry up to three attempts (30–240,s,
\texttt{tenacity}).

\subsection{Experimental Design}
\label{sec:methodology:design}

The grid comprises 18 conditions: 3 models $\times$ 3 architectures
$\times$ 2 languages (Table~\ref{tab:grid}). All conditions use the
complete dataset ($n = 625$, temperature $= 0.0$); seed $42$ only fixes the
dataset-sampling routine, which here selects every requirement.
Conditions run sequentially on a single machine to avoid GPU
contention, and the runner supports resumption via \texttt{--resume}.
The replication package (prompts, manifests, raw predictions, and
analysis scripts) is archived on Zenodo
(\url{https://doi.org/10.5281/zenodo.23380444}); the dataset is not
redistributed, and its file checksum is recorded in every manifest and in
the package notes.

\begin{table}[!ht]
\caption{Experimental grid (18 conditions).}
\label{tab:grid}
\begin{tabular}{ll}
\toprule
\textbf{Factor} & \textbf{Levels} \\
\midrule
Architecture & baseline, pipeline, two-call baseline \\
Model        & qwen2.5:7b, llama3.1:8b, mistral:7b \\
Language     & EN (original), PT (machine-translated) \\
\bottomrule
\end{tabular}
\\[2pt]
\small $n = 625$, temperature $= 0.0$, seed $= 42$
\end{table}

\subsection{Evaluation Metrics and Statistical Analysis}
\label{sec:methodology:metrics}

\textbf{Classification (RQ1, language effect).} The primary outcome is
binary FR/NFR classification against PROMISE NFR+ labels, reported as
Macro F1 (primary; it weights the F and NF classes equally despite the
255 vs.\ 370 imbalance), Accuracy, and the Matthews Correlation
Coefficient (MCC). NFR subcategory F1 is diagnostic and computed over the
ground-truth NF requirements also classified correctly as NF; subcategory
precision is computed against every emission of that label in the
condition. Categories with $n < 30$ (PO $n=1$, FT $n=10$, MN $n=17$, A
$n=21$, SC $n=21$) are excluded from the primary conclusions
(\S\ref{sec:results:nfr}).

\textbf{MoSCoW consistency (quality observation).} As no priority ground
truth exists, we measure inter-model agreement with Fleiss'
$\kappa$~\cite{fleiss1971kappa}, treating the three models as raters,
within each of the 6 (architecture $\times$ language) strata
($n_{\text{aligned}} = 623$ in all; the 2 dropped texts are the same
alignment exclusion used for RQ1), interpreted with the Landis--Koch
scale~\cite{landis1977kappa}. To test whether the pipeline's increment
over the two-call baseline is distinguishable from zero, we run a paired
percentile bootstrap ($B = 10{,}000$, same resampled indices for both
architectures) on $\Delta\kappa = \kappa_{\text{pipeline}} -
\kappa_{\text{two-call}}$, with Bonferroni correction over $N = 2$
language strata ($\alpha' = 0.025$). Because the bootstrap resamples
requirements but not executions, we also re-executed the twelve pipeline
and two-call conditions once and compare $\Delta\kappa$ across the two
executions and against the $\kappa$ variation between repeated runs
(\S\ref{sec:results:moscow}).

\textbf{RQ1 paired comparison.} For each of the 6 (model $\times$
language) pairs, the exact McNemar test~\cite{mcnemar1947note} on the
discordant pairs compares per-requirement outcomes of pipeline and
baseline over $n = 623$ aligned requirements (joined by requirement text,
since \texttt{Requirement.id} is regenerated on every load; two texts occur
twice verbatim in PROMISE NFR+ and collapse to one key each, so $n=623$;
under exclusion a pair also needs a non-fallback prediction on both sides,
giving effective $n$ of 621, 623, 623, 590, 623, and 622 for the six
conditions). Because the outcomes are binary,
the Wilcoxon signed-rank test~\cite{wilcoxon1945individual} on the same
paired differences reduces to the same discordant-pair counts; it is the primary test on the fallback-excluded data, and we report both.
Both test per-requirement correctness, not Macro-F1, so we also test the
Macro-F1 differences directly with a paired bootstrap resampling
requirements within each true class ($B = 10{,}000$; Bonferroni-adjusted
intervals, $\alpha' = 0.05/6$, over the 6 comparisons per contrast). The
effect size is Cohen's $h$~\cite{cohen1988statistical} (thresholds
$0.20/0.50/0.80$). Because $h$ uses only marginal accuracies, we also
report the paired $g = b/(b+c) - 0.5$ on the $b+c$ discordant pairs with
an exact (Clopper--Pearson) 95\% interval and thresholds
$0.05/0.15/0.25$; a magnitude is estimable only if its interval lies
within one band. Bonferroni correction over 6 comparisons gives
$\alpha' = 0.05/6 = 0.0083$.

\textbf{Language effect.} For each of the 9 (architecture $\times$ model)
pairs, PT and EN outcomes are aligned by requirement text and compared
with the exact McNemar test ($n = 623$; $\alpha' = 0.05/9 = 0.0056$). An
earlier unpaired analysis (Mann--Whitney U) is inappropriate, since both
languages classify the same requirements, and is reported only for
comparison. Uncorrected $p$-values are reported throughout.

\textbf{Parse-failure sensitivity.} The primary figures exclude the 38
confidence-zero fallbacks (\S\ref{sec:results:errorprop}). As a
sensitivity analysis we repeat the RQ1, $\Delta_{\text{p}}$ and
$\Delta_{\text{s}}$, and language-effect analyses with those fallbacks
retained as incorrect predictions and report both readings; the excluded
readings are in \texttt{statistical\_results.json}, the retained ones in
\texttt{statistical\_results\_retained.json}.

\textbf{Ablation contrasts.} $\Delta_{\text{s}}$ (pipeline vs.\ two-call)
uses the same protocol as RQ1 (Wilcoxon, Cohen's $h$, $n = 623$,
$\alpha' = 0.0083$) to test whether it is distinguishable from zero. As a
direct, non-subtractive check we also compare the baseline against the
two-call baseline, where the typed contract is absent on both sides, so a
significant difference reflects decomposition and its accompanying
prompts rather than the typed state (\S\ref{sec:results:rq2}).

\section{Results}
\label{sec:results}

We report RQ1 (\S\ref{sec:results:rq1}), the language-effect analysis
(\S\ref{sec:results:language}), MoSCoW consistency (\S\ref{sec:results:moscow}),
NFR subcategory diagnostics (\S\ref{sec:results:nfr}), RQ2
(\S\ref{sec:results:rq2}), and validation timing
(\S\ref{sec:results:errorprop}).

\textbf{In short,} most of the pipeline's F1 gain is associated with the two-call variant with
specialised prompts rather than with the typed state, which is accompanied by earlier containment of
out-of-range values and a small, unisolated consistency difference.

\textbf{Uniform exclusion policy.} All accuracy, F1, and MCC figures are
computed after excluding confidence-zero parse-failure fallbacks (38
outputs across 4 of the 18 conditions), which are infrastructure
artefacts of the shared parsing layer rather than model predictions,
applied identically to all conditions (\S\ref{sec:results:errorprop}
details which are affected). Because exclusion is an analyst's choice, we
also report the reading that \emph{retains} them as incorrect
predictions: McNemar tests and Cohen's $g$ are computed on all 623
aligned pairs, and the Wilcoxon sensitivity is stated wherever it changes
a conclusion. Exclusion is the primary analysis because it isolates model behaviour;
the retained reading is the end-to-end sensitivity analysis, in which a
parsing failure is an unsuccessful system outcome, and it is reported in
full beside it. The conclusions of 5 of the 6
RQ1 conditions and of 8 of the 9 language-effect cells are unaffected; the exceptions are discussed
in \S\ref{sec:results:rq1}--\S\ref{sec:results:rq2}.

\subsection{RQ1: Pipeline vs. Baseline Classification}
\label{sec:results:rq1}

\begin{table*}[!tb]
\caption{RQ1, pipeline vs.\ baseline (primary metric: macro F1). Acc, F1 and MCC exclude confidence-zero parse-failure fallbacks (\S\ref{sec:results:errorprop}). $p_W$: Wilcoxon signed-rank on the same data. $p_M$: exact McNemar on 623 pairs retaining the fallbacks (Bonferroni $\alpha'=0.0083$). $\checkmark$: significant in per-requirement correctness in the primary analysis (fallbacks excluded, $p_W < \alpha'$); it does not mark the macro-F1 bootstrap of Table~\ref{tab:f1boot}. $\ddagger$: significant only in the correctness test that retains fallbacks ($p_M < \alpha'$). $g$: paired Cohen's $g$; $\dagger$: interval spans several magnitude bands, not reliably estimable.}
\label{tab:rq1}
\small
\centering
\begin{tabular}{llrrrrrrrrrrrl}
\toprule
& & \multicolumn{3}{c}{Baseline} & \multicolumn{3}{c}{Pipeline} & & & & & & \\
\textbf{Model} & \textbf{L} & Acc & F1 & MCC & Acc & F1 & MCC & $\Delta$F1 & $h$ & $p_W$ & $p_M$ & $g$ & Sig. \\
\midrule
qwen2.5:7b  & PT & 0.733 & 0.733 & 0.527 & 0.814 & 0.812 & 0.638 & +0.079 & +0.193 & $3.3\times10^{-7}$ & $6.4\times10^{-7}$ & +0.253 & \checkmark \\
qwen2.5:7b  & EN & 0.784 & 0.784 & 0.613 & 0.808 & 0.807 & 0.639 & +0.023 & +0.060 & 0.059 & 0.077 & +0.119$^\dagger$ & – \\
llama3.1:8b & PT & 0.640 & 0.630 & 0.446 & 0.696 & 0.693 & 0.511 & +0.063 & +0.119 & $7.4\times10^{-6}$ & $7.7\times10^{-6}$ & +0.287 & \checkmark \\
llama3.1:8b & EN & 0.728 & 0.724 & 0.557 & 0.730 & 0.728 & 0.556 & +0.004 & $-$0.011 & 0.680 & 0.0073 & +0.154$^\dagger$ & $\checkmark^\ddagger$ \\
mistral:7b  & PT & 0.602 & 0.586 & 0.380 & 0.579 & 0.560 & 0.345 & $-$0.026 & $-$0.046 & 0.071 & 0.092 & $-$0.117$^\dagger$ & – \\
mistral:7b  & EN & 0.606 & 0.591 & 0.394 & 0.638 & 0.628 & 0.440 & +0.037 & +0.066 & 0.005 & 0.0066 & +0.200 & \checkmark \\
\bottomrule
\end{tabular}
\end{table*}

\begin{table*}[!tb]
\caption{Paired, class-stratified bootstrap of macro-F1 differences ($B=10{,}000$). Shown: point estimate and Bonferroni-adjusted interval ($\alpha'=0.05/6$, i.e.\ a 99.17\% percentile interval); bold when that interval excludes zero. $n$: requirements valid (not a fallback) for both sides of that contrast. Point estimates can differ slightly from Tables~\ref{tab:rq1} and~\ref{tab:ablation}, which compute each F1 on its own valid set. Unadjusted 95\% intervals are in \texttt{macro\_f1\_bootstrap.json}.}
\label{tab:f1boot}
\small
\centering
\begin{tabular}{llcrcrcr}
\toprule
 & & \multicolumn{2}{c}{Pipeline $-$ Baseline} & \multicolumn{2}{c}{$\Delta_{\text{p}}$ (2-call $-$ Base)} & \multicolumn{2}{c}{$\Delta_{\text{s}}$ (Pipe $-$ 2-call)} \\
\cmidrule(lr){3-4}\cmidrule(lr){5-6}\cmidrule(lr){7-8}
\textbf{Model} & \textbf{L} & $\Delta$ [adj.\ interval] & $n$ & $\Delta$ [adj.\ interval] & $n$ & $\Delta$ [adj.\ interval] & $n$ \\
\midrule
qwen2.5:7b & PT & \textbf{+0.079 [+0.042, +0.119]} & 621 & \textbf{+0.076 [+0.034, +0.115]} & 623 & +0.005 [$-$0.011, +0.020] & 621 \\
qwen2.5:7b & EN & +0.023 [$-$0.009, +0.057] & 623 & +0.020 [$-$0.013, +0.054] & 623 & +0.003 [$-$0.005, +0.013] & 623 \\
llama3.1:8b & PT & \textbf{+0.063 [+0.028, +0.100]} & 623 & \textbf{+0.053 [+0.018, +0.087]} & 623 & +0.011 [$-$0.003, +0.026] & 623 \\
llama3.1:8b & EN & $-$0.005 [$-$0.041, +0.028] & 590 & $-$0.003 [$-$0.036, +0.029] & 590 & $-$0.002 [$-$0.013, +0.010] & 623 \\
mistral:7b & PT & $-$0.027 [$-$0.066, +0.011] & 623 & $-$0.024 [$-$0.062, +0.015] & 623 & $-$0.003 [$-$0.014, +0.006] & 623 \\
mistral:7b & EN & \textbf{+0.037 [+0.003, +0.073]} & 622 & +0.024 [$-$0.008, +0.058] & 621 & \textbf{+0.013 [+0.002, +0.027]} & 621 \\
\bottomrule
\end{tabular}
\end{table*}

The pipeline achieves statistically significant improvements in 3 of 6
conditions after Bonferroni correction under the primary (exclusion)
analysis, and in 4 of 6 under the paired McNemar test that retains the
fallbacks (Table~\ref{tab:rq1}). These tests compare per-requirement correctness; a
paired, class-stratified bootstrap of the macro-F1 difference
(Table~\ref{tab:f1boot}) gives the same verdict under exclusion: the
Bonferroni-adjusted interval excludes zero for \texttt{qwen2.5:7b} PT,
\texttt{llama3.1:8b} PT, and \texttt{mistral:7b} EN, and for
$\Delta_{\text{p}}$ in the first two. For $\Delta_{\text{s}}$ it
excludes zero only for \texttt{mistral:7b} EN. The fourth is
\texttt{llama3.1:8b} EN,
non-significant under exclusion ($p = 0.680$) but significant under
McNemar ($p = 0.0073$) and under Wilcoxon with fallbacks retained
($p = 0.0055$): the 33 fallbacks removed from the baseline in that cell
hid the gap. Cohen's $h$ places every effect in the negligible range
($|h| < 0.20$), but $h$ ignores the pairing. The paired $g$ is large for
\texttt{qwen2.5:7b} PT and \texttt{llama3.1:8b} PT and medium for
\texttt{llama3.1:8b} EN and \texttt{mistral:7b} EN, and its 95\% interval
stays within one magnitude band only for the first two and for
\texttt{mistral:7b} EN; for the other three, the $50$--$97$ discordant
pairs do not support a confident magnitude. We therefore do not claim
that all effects are negligible: they are modest by $h$ and
non-negligible by $g$ where $g$ is estimable. The largest gain is
\texttt{qwen2.5:7b} PT ($\Delta = +7.9$\,pp, $g = +0.253$, F1
$0.812$ vs.\ $0.733$).

\texttt{mistral:7b} PT is the only case where the pipeline leans toward
underperformance ($\Delta = -2.6$\,pp, $p = 0.071$), consistent with a
capacity-floor hypothesis. The zero-confidence count does not support
it (\texttt{mistral:7b} has at most 2/625 such fallbacks in any
condition), so we use a finer compliance proxy: the share of predicted-NF outputs without a valid
\texttt{nfr\_category} is 7.3\% (pipeline) and 8.1\% (two-call) for
\texttt{mistral:7b} PT, against 1.3--1.4\% in English and 0.3--2.6\% in
the other cells with a non-zero rate, i.e.\ about 5--6$\times$ its
English rate (\S\ref{sec:results:errorprop}). We present the capacity
floor as a hypothesis, not a tested mechanism.

\textbf{Answer to RQ1:} On requirements valid for both sides, the paired
bootstrap supports a Macro-F1 difference in favour of the pipeline in 3
of 6 conditions (Bonferroni-adjusted). In per-requirement correctness,
the Wilcoxon test agrees in the same 3, and with the fallbacks retained
the McNemar test finds a difference in 4 of 6. Effect sizes are modest by
Cohen's $h$ but medium to large by the paired $g$ where estimable. The
difference depends on the model's ability to produce structured output;
models below a compliance threshold show no gain from the specialised
two-call pipeline.

\subsection{Complementary Analysis: PT vs.\ EN Conditions}
\label{sec:results:language}

Because PT and EN classify the \emph{same} requirements, the paired
analysis is the appropriate one: aligning them by requirement text and
applying the exact McNemar test (column $p_M$ of Table~\ref{tab:rq2};
$n=623$, fallbacks retained, $\alpha' = 0.0056$) shows a
significant PT$<$EN gap in five of nine cells: the baseline with
\texttt{qwen2.5:7b} and \texttt{llama3.1:8b}, the pipeline with
\texttt{mistral:7b}, and the two-call baseline with
\texttt{llama3.1:8b} and \texttt{mistral:7b}. English is the better
language in every significant cell, and only \texttt{qwen2.5:7b} is
language-insensitive in both decomposed architectures ($p = 0.91$ and
$0.82$). An earlier unpaired Mann--Whitney analysis, kept in
Table~\ref{tab:rq2} for comparison only, reached significance after
correction only for the baseline with \texttt{llama3.1:8b}
($h = -0.191$, $p < 0.001$). That baseline \texttt{llama3.1:8b} result also depends
on the exclusion policy: with the 33 EN fallbacks retained, the
Mann-Whitney test gives $p = 0.063$ and no longer survives correction,
while the paired test is significant ($p = 0.0009$).

\begin{table}[!ht]
\caption{Language-effect results. $p_{MW}$: Mann-Whitney U (unpaired, original analysis; fallbacks excluded). $p_M$: exact McNemar on 623 PT/EN-aligned pairs (fallbacks retained); bold = significant after Bonferroni over 9 comparisons ($\alpha'=0.0056$). Accuracies follow the exclusion policy of \S\ref{sec:results}; four of the 18 cells are affected (\S\ref{sec:results:errorprop}).}
\label{tab:rq2}
\resizebox{\linewidth}{!}{%
\begin{tabular}{llrrrrr}
\toprule
\textbf{Arch.} & \textbf{Model} & PT & EN & $h$ & $p_{MW}$ & $p_M$ \\
\midrule
Baseline & qwen2.5:7b  & 0.734 & 0.785 & $-$0.120 & 0.034 & \textbf{$3.0\times10^{-5}$} \\
Baseline & llama3.1:8b & 0.640 & 0.729 & $-$0.191 & $<0.001$ & \textbf{0.0009} \\
Baseline & mistral:7b  & 0.602 & 0.607 & $-$0.010 & 0.862 & 0.801 \\
Pipeline & qwen2.5:7b  & 0.815 & 0.809 & +0.015 & 0.793 & 0.913 \\
Pipeline & llama3.1:8b & 0.697 & 0.730 & $-$0.075 & 0.188 & 0.011 \\
Pipeline & mistral:7b  & 0.579 & 0.638 & $-$0.121 & 0.034 & \textbf{$9.1\times10^{-6}$} \\
Two-call & qwen2.5:7b  & 0.811 & 0.806 & +0.012 & 0.829 & 0.824 \\
Two-call & llama3.1:8b & 0.687 & 0.732 & $-$0.099 & 0.081 & \textbf{0.0004} \\
Two-call & mistral:7b  & 0.583 & 0.626 & $-$0.090 & 0.115 & \textbf{0.0004} \\
\bottomrule
\end{tabular}%
}
\end{table}

\textbf{Language-effect finding:} Under the paired analysis, Portuguese
is associated with significantly lower accuracy than English in 5 of 9
architecture--model cells, including decomposed architectures for
\texttt{mistral:7b} and \texttt{llama3.1:8b}. We therefore do not claim
that decomposition attenuates the language effect, nor present the
results as evidence of multilingual portability. The data support only
that \texttt{qwen2.5:7b} is language-insensitive under both decomposed
architectures and that the baseline \texttt{llama3.1:8b} gap is not
specific to that architecture. Because requirement and prompt language
change together, and the Portuguese text is a machine translation, the
gap cannot be attributed to either (\S\ref{sec:discussion:external}).

\subsection{MoSCoW Consistency: A Software Quality Observation}
\label{sec:results:moscow}

Because PROMISE NFR+ provides no MoSCoW ground truth, we assess
prioritisation as \textbf{inter-model behavioural consistency},
operationalised as Fleiss' $\kappa$ over the three models within each
(architecture $\times$ language) stratum (Table~\ref{tab:kappa}). It is a
potential proxy for the \textit{Reliability} characteristic of
\S\ref{sec:background}, not a measure of prioritisation quality: models
may agree and still be wrong, and we have neither priority ground truth
nor practitioner evaluation.

\begin{table}[!ht]
\caption{Inter-model MoSCoW consistency, Fleiss' $\kappa$ (3 raters,
6 strata), full four-label priority set. $n$: requirements aligned
across all three raters (same filter as RQ1).}
\label{tab:kappa}
\begin{tabular}{llrrl}
\toprule
\textbf{Architecture} & \textbf{L} & $n$ & $\kappa$ & Interp. \\
\midrule
Baseline  & PT & 623 & $-$0.1406 & poor \\
Baseline  & EN & 623 & $-$0.1220 & poor \\
Two-call  & PT & 623 & +0.1360 & slight \\
Two-call  & EN & 623 & +0.1695 & slight \\
Pipeline  & PT & 623 & +0.1768 & slight \\
Pipeline  & EN & 623 & +0.2388 & fair \\
\bottomrule
\end{tabular}
\end{table}

The negative $\kappa$ in both baseline strata reflects systematic
strategy divergence: in the PT baseline, \texttt{llama3.1:8b} and
\texttt{mistral:7b} collapse to a Must rate of 99\% and 97\% while
\texttt{qwen2.5:7b} spreads evenly (46\%), a configuration Fleiss'
$\kappa$ scores as random disagreement but which shows \emph{which}
models fail to discriminate. The Must rate of those two models falls to
0.90/0.86 (two-call) and 0.87/0.85 (pipeline), while \texttt{qwen2.5:7b}
stays at 0.41--0.42; the whole progression is driven by the two weaker
models reducing their Must-collapse at each step.

The six strata show a monotone progression: baseline ($\kappa \leq
-0.12$) $\to$ two-call ($\approx +0.15$) $\to$ pipeline ($\leq +0.24$).
The two-call baseline, with the pipeline's prompts but no typed state,
already recovers most of the improvement; the pipeline adds
$\Delta\kappa = +0.04$ to $+0.07$. Unlike $\Delta_{\text{s}}$ for F1,
this increment is distinguishable from resampling noise: a paired
percentile bootstrap ($B = 10{,}000$, Bonferroni over $N = 2$,
$\alpha' = 0.025$; \S\ref{sec:methodology:metrics}) gives PT
$\Delta\kappa = +0.0408$, 97.5\% CI $[+0.0133, +0.0703]$, $p = 0.0004$,
and EN $+0.0694$, CI $[+0.0310, +0.1083]$, $p < 0.0001$. This holds only
within this three-model panel (\S\ref{sec:discussion:external}). The
primary driver of the $\kappa$ gain is thus decomposition with its
task-specific prompts, matching the F1 ablation
(\S\ref{sec:results:rq2}).

The residual pipeline-vs-two-call difference is reproducible: a second
execution of the twelve pipeline and two-call conditions (same code,
seed, and configuration; the model server was updated in between,
\S\ref{sec:discussion:internal}) reproduced the sign of $\Delta\kappa$
(PT $+0.022$, EN $+0.079$) and of all four pairings of the two
executions (PT $+0.022$ to $+0.041$, EN $+0.052$ to $+0.097$), each
exceeding the $\kappa$ variation between repeated runs (at most $0.011$
in PT and $0.027$ in EN). With two executions the noise estimate is
coarse, and the PT increment roughly halved (item-bootstrap 95\% interval
$[+0.003, +0.041]$), so we regard it as a reproducible association, not
a precisely estimated effect. We do not attribute it to the typed
contract: the validation boundary rejected no real output, confidence
routing affected at most three of 3,750 prioritisations
(\S\ref{sec:architecture:limits}), and the two architectures differ on
0.6--2.1\% of types and up to 13\% of priorities, which points to
execution-order and run-to-run effects that a single-run design cannot
separate from the contract. We did not test whether the increment
translates into an operational benefit.

\subsection{NFR Subcategory F1 (Diagnostic)}
\label{sec:results:nfr}

Subcategory recall is computed over the 272 (of 370) NF requirements
also classified correctly as NF, and precision against all 625
predictions, so a subcategory label emitted on a requirement misclassified
as FR counts against that label's precision
(\S\ref{sec:methodology:metrics}). In the best condition
(\texttt{qwen2.5:7b} pipeline EN), among categories with $n \geq 30$,
SE achieves the highest F1 ($0.916$; $P = 0.927$, $R = 0.905$, $n=42$),
followed by PE ($0.701$) and LF ($0.566$), while O ($0.333$) and US
($0.429$; $P = 1.000$, $R = 0.273$) are weakest. The US imbalance
indicates a conservative model that labels Usability only when confident,
consistently across architectures, so it reflects the model's
representation of the class rather than an effect of decomposition.
These patterns do not affect the RQ1 and ablation conclusions (binary
macro F1) but mark a boundary for our quality claims: validation timing
(\S\ref{sec:results:errorprop}) targets \emph{structural} conformance,
not \emph{semantic} correctness such as category ambiguity. Per-category values (all
conditions) and the Must-label rates per model are in
\texttt{subcategory\_errorprop.json} and \texttt{kappa\_table.json}.

\subsection{RQ2: Decomposition vs.\ Typed-State Quality Assurance}
\label{sec:results:rq2}

\begin{table}[!ht]
\caption{What each contrast changes and permits us to conclude.}
\label{tab:contrasts}
\small
\begin{tabular}{@{}p{1.6cm}p{2.7cm}p{2.9cm}@{}}
\toprule
\textbf{Contrast} & \textbf{What changes} & \textbf{What it supports} \\
\midrule
$\Delta_{\text{p}}$: baseline $\to$ two-call & calls, task prompts, staged context & F1 associated with the variant (not which element) \\
$\Delta_{\text{s}}$: two-call $\to$ pipeline & validation boundary, confidence, LangGraph & bound on the F1 difference; QA properties (not their cause) \\
\bottomrule
\end{tabular}
\end{table}

Table~\ref{tab:ablation} reports the three-way comparison.
$\Delta_{\text{p}}$ is a difference associated with the two-call
specialised-prompt variant: its second call receives the already-decided
\texttt{requirement\_type} and \texttt{nfr\_category}, so it conflates
prompt wording, an additional call, and staged context (a call-count
control is future work). Read with that caveat, it is substantial in 4 of
6 conditions (up to $+0.075$ F1), so most of the pipeline's accuracy advantage over the
single-call baseline is associated with decomposition. Under
exclusion it is negligible for \texttt{llama3.1:8b} EN ($+0.006$), but
retaining the baseline's 33 fallbacks raises it to $+0.045$; it is
negative for \texttt{mistral:7b} PT, and largest for \texttt{qwen2.5:7b}
PT, the strongest instruction follower. $\Delta_{\text{s}}$ is the
residual F1 difference between the typed pipeline and the two-call
baseline; it bundles the earlier validation point, execution-order
effects, and confidence routing (triggered on only three
prioritisations, \S\ref{sec:architecture:limits}). It is small in all
conditions ($|\Delta_{\text{s}}| \leq 0.013$), inconsistent in direction,
and essentially unchanged when fallbacks are retained (largest shift
$0.003$).

\begin{table*}[!t]
\caption{Decomposition ablation (macro F1). $\Delta_p$ = F1(two-call) $-$ F1(baseline), a difference associated with the two-call variant (prompt wording, extra call, and staged context are confounded); $\Delta_s$ = F1(pipeline) $-$ F1(two-call). \emph{Excluded} drops the confidence-zero fallbacks (\S\ref{sec:results:errorprop}); \emph{Retained} keeps them as errors ($n=625$; $\Delta$ computed from unrounded F1). $b/c$: requirements correct only for the pipeline / only for the two-call baseline. $g$: paired Cohen's $g$ for $\Delta_s$; $\dagger$: not reliably estimable (4--15 discordant pairs).}
\label{tab:ablation}
\small
\centering
\begin{tabular}{llrrrrrrrrr}
\toprule
 & & & & & \multicolumn{2}{c}{Excluded} & \multicolumn{2}{c}{Retained} & & \\
\cmidrule(lr){6-7}\cmidrule(lr){8-9}
\textbf{Model} & \textbf{L} & Base & 2Call & Pipe & $\Delta_p$ & $\Delta_s$ & $\Delta_p$ & $\Delta_s$ & $b/c$ & $g$ \\
\midrule
qwen2.5:7b  & PT & 0.733 & 0.808 & 0.812 & +0.075 & +0.004 & +0.075 & +0.001 & 8/7 & +0.033$^\dagger$ \\
qwen2.5:7b  & EN & 0.784 & 0.804 & 0.807 & +0.020 & +0.003 & +0.020 & +0.003 & 3/1 & +0.250$^\dagger$ \\
llama3.1:8b & PT & 0.630 & 0.682 & 0.693 & +0.053 & +0.011 & +0.053 & +0.011 & 8/2 & +0.300$^\dagger$ \\
llama3.1:8b & EN & 0.724 & 0.730 & 0.728 & +0.006 & $-$0.002 & +0.045 & $-$0.002 & 3/4 & $-$0.071$^\dagger$ \\
mistral:7b  & PT & 0.586 & 0.563 & 0.560 & $-$0.023 & $-$0.003 & $-$0.023 & $-$0.003 & 1/3 & $-$0.250$^\dagger$ \\
mistral:7b  & EN & 0.591 & 0.614 & 0.628 & +0.023 & +0.013 & +0.024 & +0.013 & 7/0 & +0.500 \\
\bottomrule
\end{tabular}
\end{table*}

On per-requirement correctness, a paired Wilcoxon test (Bonferroni
$\alpha' = 0.0083$) finds five of six $\Delta_{\text{s}}$ comparisons not
significant; the sixth (\texttt{mistral:7b} EN, $p = 0.0082$) narrowly
reaches significance despite a small effect ($h = +0.023$). The
Macro-F1 bootstrap of Table~\ref{tab:f1boot} agrees: only that
condition has an adjusted interval excluding zero. The pipeline and the two-call
baseline disagree on only 4--15 requirements per condition, so the
interval of $g$ spans several magnitude bands in five of six conditions;
only \texttt{mistral:7b} EN ($7$ vs.\ $0$, $g=+0.50$, $[+0.09, +0.50]$)
is reliably non-negligible. We therefore report these effects as small in
F1 and not distinguishable, at this $n$, from effects of the size the
interval allows.

A small $\Delta_{\text{s}}$ does not show that the typed contract has no
quality value; it shows that F1 registers no difference, while other
instruments do (\S\ref{sec:results:moscow} for $\kappa$ and
\S\ref{sec:results:errorprop} for validation timing). It does not
establish that the typed contract causes those differences.

As a direct, non-subtractive check, we compare the single-call baseline
with the two-call baseline head-to-head, so that the typed contract is
absent on both sides (same protocol as RQ1, $\alpha' = 0.0083$).
Decomposition alone is significant in 2 of 6 conditions with fallbacks
excluded (\texttt{qwen2.5:7b} PT, $h=+0.184$, $g=+0.250$;
\texttt{llama3.1:8b} PT, $h=+0.099$, $g=+0.254$; both negligible by $h$
and large by $g$), and in 3 of 6 when the 33 baseline fallbacks are
retained (\texttt{llama3.1:8b} EN, $h=+0.092$, McNemar $p=0.0038$), the
same flip that RQ1 shows. This reproduces the RQ1 pattern without the
subtraction logic of $\Delta_{\text{p}}$ and $\Delta_{\text{s}}$,
strengthening the case that where the pipeline beats the baseline, the
gain is associated with decomposition and its prompts rather than with
the typed contract.

\textbf{Answer to RQ2:} Most of the F1 difference between the pipeline and the
single-agent baseline is associated with decomposition and its
task-specific prompts ($\Delta_{\text{p}} \in [+0.020, +0.075]$ in 4 of 6
conditions), corroborated by the direct baseline-vs-two-call comparison;
we cannot separate prompt wording from the extra call and staged context.
The pipeline-minus-two-call difference is small and directionally inconsistent
($\Delta_{\text{s}} \in [-0.003, +0.013]$), and the discordant pairs are
too few to bound its magnitude. Two narrow pieces of evidence concern the
typed contract's quality-assurance properties: earlier structural
validation (\S\ref{sec:results:errorprop}, 13 injected faults) and a
statistically confirmed, reproducible difference in inter-model MoSCoW
consistency (\S\ref{sec:results:moscow}) whose cause the design does not
isolate.

\textbf{Supplementary check with a larger model.} To test whether the
pipeline's gain depends on the small models' compliance, we ran the three
architectures on \texttt{gpt-4.1-mini} (Azure AI Foundry), outside the
18-condition grid, with the same prompts, seed, and temperature; no
classification output fell back. The pipeline outperforms the baseline in
accuracy (PT $+0.019$, McNemar $p=0.043$; EN $+0.034$, $p<10^{-4}$,
$n=623$; only EN survives Bonferroni over two languages,
$\alpha'=0.025$). The gain is associated with decomposition: $\Delta_{\text{p}}$ is
$+0.013$ (PT, $p=0.17$) and $+0.037$ (EN, $p<10^{-4}$), whereas
$\Delta_{\text{s}}$ is $+0.006$ (PT, $p=0.42$) and $-0.003$ (EN,
$p=0.63$), reproducing the small-model pattern. Caveats: one model, one
execution per condition; the PT baseline and pipeline and both two-call
runs were executed on an adjacent development line whose agents differ in
category normalisation, JSON extraction, and confidence forwarding (none
affects the binary accuracy reported, as no classification fallbacks
occurred), and only the EN baseline and pipeline were re-executed with the
artifact's agents; and 32 pipeline and 37 two-call English
prioritisations (5--6\%) fell back to Could Have because of a trailing
comma in the model's JSON, which does not enter the accuracy figures but
would affect MoSCoW analyses.

\subsection{Validation Timing and Schema Compliance}
\label{sec:results:errorprop}

The clearest evidence concerning the typed state's quality-assurance
properties is a difference in \emph{when} structural validation occurs,
not a metric. In all three architectures, a parse or enum-conversion
failure is caught inside the responsible agent's \texttt{\_parse\_response}
and replaced with a conservative default (\texttt{FUNCTIONAL},
\texttt{confidence=0.0}); no malformed value reaches a downstream agent's
context. This robustness is not attributable to the typed contract: all
four agents use the same regex-based JSON extraction and fallback,
duplicated across the codebase.

The typed contract's verifiable contribution is validation \emph{timing}.
In the pipeline, \texttt{ClassificationOutput}'s Pydantic constraints
(e.g., \texttt{confidence} in $[0,1]$) are enforced immediately after the
\texttt{ClassificationAgent} call, before the result enters
\texttt{PipelineState}. In the two-call ablation the intermediate is a
plain \texttt{@dataclass}, and Pydantic validation occurs only when the
final output is assembled after the second call. The typed contract thus
validates one LLM call earlier, a narrow V\&V property: a violation could
be acted on before the second, more expensive call.

On the PROMISE NFR+ grid this difference never surfaced as a failure: no
confidence or type value fell outside its bounds in any of the 18 runs.
Parse-failure fallbacks are rare: 38 outputs across 4 runs (33 baseline
\texttt{llama3.1:8b} EN, 5.3\%; 2 pipeline \texttt{qwen2.5:7b} PT; 1
pipeline \texttt{mistral:7b} EN; 2 two-call \texttt{mistral:7b} EN). We
exclude them from Tables~\ref{tab:rq1}, \ref{tab:rq2}, and
\ref{tab:ablation} and report the retained reading as a sensitivity
analysis (\S\ref{sec:results}). The zero-confidence count is a poor proxy
for structured-output compliance; a finer one is the share of predicted-NF
outputs with no valid \texttt{nfr\_category}: for \texttt{mistral:7b} it
is 7.3\% (pipeline) and 8.1\% (two-call) in Portuguese against 1.3--1.4\%
in English, and no other condition exceeds 2.6\%, which is what the
capacity-floor reading of \S\ref{sec:results:rq1} relies on. The three
pipeline fallbacks are the only outputs with confidence below $0.70$, so
only 3 of 3{,}750 prioritisations received the conservative-label
instruction; routing was effectively inert, and no reported result
depends on it.

To verify the boundary independently of the grid, we ran a systematic
fault-injection campaign through the real agents of each architecture with
a scripted LLM (\texttt{run\_fault\_injection\_campaign.py}). It injects 13
faults into the model's JSON, in either call: six value-range violations
(\texttt{confidence} $1.4$, $-0.5$, NaN; \texttt{priority\_score} $1.5$,
$-0.1$; \texttt{priority\_rank} $0$) and seven parse-level faults (invalid
or missing enumeration, wrong type, truncated JSON). The typed contract adds nothing for parse-level faults, which the
two-call baseline and the pipeline handle identically (a flagged fallback
after two calls). The
difference lies in the value-range violations (Table~\ref{tab:faults}). The
typed pipeline rejects each where the typed object is built, detecting
classification-stage faults after the first call and prioritisation-stage
faults after the second, and keeps the requirement as a flagged fallback
after two calls. The unvalidated two-call baseline passes the same values
through its dataclass and fails when the output model is assembled, after
the second call; because that failure propagates through the retry
wrapper, both calls are repeated up to three attempts, consuming six calls
before the requirement is dropped without a failure record. The
single-call baseline contains all 13 faults at its one call, and no
architecture let an out-of-range value reach the output. The campaign is
synthetic: it characterises where each architecture detects a fault and
what that costs, not how often such faults occur; that earlier
validation prevents an operational failure remains an untested
hypothesis. The six-call cost and
the loss in the unvalidated arm partly reflect the shared retry policy, which
retries a deterministic validation error, so the contrast is between a
contained and an uncontained failure rather than between detection and
non-detection.

\begin{table}[!ht]
\caption{Fault-injection campaign: outcome and LLM calls consumed per fault (13 faults: 6 value-range, 7 parse-level).}
\label{tab:faults}
\small
\begin{tabular}{lll}
\toprule
\textbf{Architecture} & \textbf{Value-range (6)} & \textbf{Parse-level (7)} \\
\midrule
Baseline  & flagged, 1 call & flagged, 1 call \\
Two-call  & dropped, 6 calls & flagged, 2 calls \\
Pipeline  & flagged, 2 calls & flagged, 2 calls \\
\bottomrule
\end{tabular}
\end{table}

Two conclusions follow. Structured-JSON compliance is achieved
independently in each architecture, not through one centralised
component. And validation \emph{timing}, not failure containment, is what
distinguishes the typed pipeline from the unvalidated ablation; the isolated
effect of confidence routing could not be evaluated, since it fired on
only three fallbacks.

\section{Discussion and Threats to Validity}
\label{sec:discussion}

The experiments indicate that the typed contract is accompanied by
little difference in classification performance
($\Delta_{\text{s}} \in [-0.003, +0.013]$). The evidence for what it is
accompanied by is limited to two properties: earlier structural
validation, verified by a 13-fault injection campaign, and a
statistically confirmed difference in inter-model MoSCoW consistency
whose cause we did not isolate. We do not claim, because we did not test,
benefits such as predictable behaviour under model substitution or
auditable intermediate decisions, and even these would depend on the
model satisfying the structured-output contract. For \texttt{mistral:7b}
PT (\S\ref{sec:results:rq1}, \S\ref{sec:results:errorprop}), parse
failures fall back to \texttt{FUNCTIONAL} rather than NF, silently
biasing misclassifications toward the functional class; since the model
has no more zero-confidence outputs than the others but 5--6$\times$ more
NF predictions without a valid category in Portuguese, these failures
are subtler than malformed responses. Structured-output compliance is
therefore a prerequisite for benefiting from decomposition.

Practically, teams interested mainly in F1 should invest in better
prompts rather than a typed contract, whereas applications that value
earlier structural validation or cross-model consistency may prefer the
pipeline despite its limited accuracy gain over the two-call baseline.
Our paired analysis does not support deploying the same architecture
bilingually without adaptation (\S\ref{sec:results:language}). Although the
pipeline makes two sequential calls, local execution times were
comparable to the baseline (\texttt{qwen2.5:7b} PT: 3529\,s vs.\ 3415\,s
for $n=625$); hosted APIs may approach a $2\times$ wall-clock cost, which
we did not test.

\label{sec:discussion:internal}
\textbf{Internal threats.} PROMISE NFR+ has no MoSCoW ground truth, so
prioritisation quality is operationalised as inter-model stability,
reflecting reliability rather than correctness; some requirements also
admit several plausible NFR categories while the dataset gives one label.
The principal threat is the prompt--architecture confound: the baseline
combines both subtasks in one prompt, so $\Delta_{\text{p}}$ is a
difference associated with the two-call variant, bundling prompt wording,
an extra call, and staged context, and a call-count-controlled arm would be needed to separate
them. $\Delta_{\text{s}}$ likewise bundles the validation boundary,
confidence handling (which fired for 3 of 3{,}750 predictions), and
LangGraph orchestration, so it is a pipeline-versus-two-call contrast
rather than an isolated typed-contract effect, and its range is an upper
bound on the F1 difference associated with the typed contract. Cohen's
$h$ places all RQ1 effects below $0.20$ but ignores the pairing; the
paired $g$ is medium or large where estimable, so we do not call the
effects negligible. For $\Delta_{\text{s}}$ the 4--15 discordant pairs
are too few for any reliable paired effect size. Excluding 38 fallbacks
is an analyst's decision, hence both readings; it changes one RQ1 verdict
(\texttt{llama3.1:8b} EN) and one $\Delta_{\text{p}}$ estimate. Decoding
at temperature~0 is not bit-deterministic: primary results come from one
execution per condition, and the one repeated execution (pipeline and
two-call) ran on a newer Ollama release (0.35.1 vs.\ 0.24.0), so its
variation mixes run-to-run and server-version effects; differences below
about one point of F1 (and $0.03$ in $\kappa$) should not be interpreted.
The fault-injection evidence rests on 13 synthetic faults (six of them
value-range violations, which extend the earlier five-value probe): it
shows where validation occurs and
what it costs, not how often such faults arise. Per-call traces
were retained only for the two-call supplement runs, which limits audit of
individual failures.

\label{sec:discussion:external}
\textbf{External threats.} All experiments use PROMISE NFR+ and three
decoder-only SLMs (7--8B parameters) served through versioned Ollama
tags, whose behaviour may change; compliance rates and the contribution
of typed contracts may differ for other datasets, domains, or models, and
the $\kappa$ gain reflects only this three-model panel. The single larger
model (\S\ref{sec:results:rq2}) is one model with one execution per
condition. The language comparison varies requirement and prompt language
jointly, and the Portuguese text is a machine translation (back-translation
BERTScore F1 0.9646; 624 of 625 pairs $\geq 0.90$), so translation
artefacts may contribute to the PT--EN gap. Only the pipeline used
LangGraph, although the typed-state pattern is not specific to it.
Bonferroni correction was applied throughout, and categories with fewer
than 30 instances were excluded from the primary analyses. Our
conclusions are therefore intentionally narrow: for the evaluated
dataset, task, and model panel, the measurable differences that accompany
the typed contract are earlier structural validation and a small
difference in inter-model consistency, rather than improvements in
classification accuracy.

\section{Conclusion and Future Work}
\label{sec:conclusion}

\textbf{In the evaluated setting, the accuracy difference is associated with decomposition into
specialised prompts, not with the typed state} (\S\ref{sec:results:rq1}--\S\ref{sec:results:rq2}).
The typed state is accompanied by two differences beyond accuracy: containment of out-of-range values
at the agent boundary, with classification faults detected one call earlier in a 13-fault campaign
(\S\ref{sec:results:errorprop}), and a small, reproducible increase in inter-model MoSCoW
consistency (paired bootstrap, $p < 0.001$ in both languages, \S\ref{sec:results:moscow}). The
design does not isolate their causes, and the limits are detailed in \S\ref{sec:discussion}.

\textbf{Future Work.} (i) Calibrating the confidence threshold, to
isolate routing from validation timing. (ii) A call-count-controlled arm
to separate prompt wording from the extra call and staged context.
(iii) Larger models and a larger rater panel, to test whether
$\Delta_{\text{s}}$ grows with capacity and the consistency gain
generalises.

\raggedbottom
\section*{ARTIFACT AVAILABILITY}
The implementation, prompts, run manifests, raw predictions, analysis scripts, and unit
tests are archived on Zenodo (\url{https://doi.org/10.5281/zenodo.23380444}). The PROMISE
NFR+ file is not redistributed; its MD5 hash is recorded in every manifest.

\ifanonymous\else
\section*{ACKNOWLEDGMENTS}
For partially supporting this work, we would like to thank INES.IA (National
Institute of Science and Technology for Software Engineering Based on and for
Artificial Intelligence), \url{https://www.ines.org.br}, CNPq grant
408817/2024-0.
\fi

\bibliographystyle{ACM-Reference-Format}
\bibliography{referencias}

\end{document}